# Controlling electrostatic co-assembly using ion-containing copolymers : from surfactants to nanoparticles

### J.-F. Berret

Matière et Systèmes Complexes, UMR 7057 CNRS Université Denis Diderot Paris-VII, Bâtiment Condorcet, 10 rue Alice Domon et Léonie Duquet, 75205 Paris (France)

**Abstract :** In this review, we address the issue of the electrostatic complexation between charged-neutral diblock copolymers and oppositely charged nanocolloids. We show that nanocolloids such as surfactant micelles and iron oxide magnetic nanoparticles share similar properties when mixed with charged-neutral diblocks. Above a critical charge ratio, core-shell hierarchical structures form spontaneously under direct mixing conditions. The core-shell structures are identified by a combination of small-angle scattering techniques and transmission electron microscopy. The formation of multi-level objects is driven by the electrostatic attraction between opposite charges and by the release of the condensed counterions. Alternative mixing processes inspired from molecular biology are also described. The protocols applied here consist in screening the electrostatic interactions of the mixed dispersions, and then removing the salt progressively as example by dialysis. With these techniques, the oppositely charged species are intimately mixed before they can interact, and their association is monitored by the desalting kinetics. As a result, sphere- and rod-like aggregates with remarkable superparamagnetic and stability properties are obtained. These findings are discussed in the light of a new paradigm which deals with the possibility to use inorganic nanoparticles as building blocks for the design and fabrication of supracolloidal assemblies with enhanced functionalities.

**outline**
I - Introduction
II – Electrostatic Complexation between oppositely charged surfactants and copolymers
        II.1 – Homopolyelectrolytes and surfactants
        II.2 – Evidence of critical mixing ratios
        II.3 – Hierarchical structures revealed by neutron scattering
III – Generalization of the complexation protocols to inorganic nanoparticles
        III.1 – Existing clustering strategies for inorganic nanoparticles
        III.2 – Iron oxide nanoparticles as a model system
        III.3 – Iron oxide supracolloidal assembly
IV – Desalting transition : towards the fabrication of controlled size clusters and rods
        IV.1 – Alternative protocols for mixing species with opposite charges
        IV.2 – Evidence of a desalting transition as a function of ionic strength
        IV.3 – Magnetic clusters and rods
V – Conclusion

# 1 - Introduction

The emergence of novel materials and processing at the nanoscale has set the conditions for the fabrication of a wide range of nano-objects and multilevel nanostructured networks [1-3]. In particular, the co-assembly of polymers and nanoparticles is opening pathways for engineering versatile hybrid structures combining the advantageous properties of both the organic and inorganic worlds. Interest stems from the combination of complementary attributes, such as a size in the nanometer range and unique physical features including high reactivity or magnetic and optical properties. With respect to assembly mechanisms, electrostatic complexation between oppositely charged macromolecular objects has gained special interest over the last years [4]. Electrostatic complexation designates the process by which co-assembly is driven by the pairing of electric charges located e.g. at the surfaces of particles or along the backbones of polymers.

This research was motivated by the fact that electrostatics is at the origin of many non-specific associations relevant in biological systems, such as protein complexes or chromatin. The chromatin is described as the lowest hierarchical level of chromosomes in eucaryotic cells. The basic repeating unit of chromatin fibers, the nucleosomes can be reconstituted *in vitro* from its pure components, mainly histones and DNA by various techniques [5]. This reconstitution process can be seen as the electrostatic complexation between an excess of polyanions, here 146 base pairs of DNA per nucleosome and a limited set of polycations, *i.e.* an octamer of cationic histones [6, 7].

A second prominent application of electrostatic complexation concerns polyelectrolyte multilayers [8, 9]. The layer-by-layer deposition of oppositely charged polyelectrolytes is a remarkable illustration of a supramolecular assembly at interfaces. Using this interfacial complexation approach, researchers have generated multicomponent or multilayer assemblies composed of different charged macromolecules [10, 11] and/or inorganic nano-objects such as particles, tubes, proteins [12, 13] resulting in thin films with high capacitance property [14] or coatings with antireflection and self-cleaning properties [15, 16].

A third class of systems where electrostatic complexation is at stake is that of charged-neutral block copolymers, also called double hydrophylic block copolymers. Unlike amphiphilic copolymers, charged-neutral blocks are soluble in aqueous media, but co-assemble in the presence of oppositely charged species. These copolymers were popularized in the mid 1990's by the group of Kataoka who was the first to suggest the formation of polyion complex micelles from a pair of oppositely charged block polypeptides with poly(ethylene glycol) segments [17-20]. Since this first attempt, it was recognized that the attractive interactions between copolymers and oppositely charged species generate new type of colloids [4, 21]. These colloids form spontaneously under direct mixing and exhibit a core-shell microstructure. The formation of mixed aggregates is generally described as the result of a nucleation and growth mechanism of a microphase made from the oppositely charged constituents [22]. Owing to the hindrance barrier represented by the shell, the growth is arrested at a size which is fixed by the dimension of the polymer. The specimens examined so far comprise synthetic [23-27] and biological [18, 28] macromolecules, multivalent counterions [29, 30], surfactant micelles [31-33, 22, 34-39]. Electrostatic complexation is also utilized in numerous applications such as formulation of personal care products, water treatment and filtration, coating, drug delivery, bioelectronics and nanocomposite technologies [4, 21].

Concerning the complexation mechanism, it is now accepted that the driving forces for association are both enthalpic and entropic in origin [40-43, 9]. This is true for homopolyelectrolytes as well as for the aforementioned charged-neutral block polymers. The enthalpic part in the free energy of association is linked to the pairing of the opposite charges. The binding enthalpy depends strongly on the chemical constituents that are paired in the complexation [9]. The entropic contribution to the free energy arises from the release of the counterions which are condensed on the surface of the colloid or along the backbone of the polymer [44], as well as from the loss of translational and rotational degrees of freedom of macromolecules in their bound state [45]. The balance between the enthalpic and entropic contributions gives rise to wide variety of complexation behaviors. The association can be strong, and then results in the formation of coacervates and macroscopic phase separations [46-48, 37]. In some systems, the complexation with ion-containing chains is weaker and the electrostatic complexes remain soluble [49, 50, 43, 51-53]. For electrostatic complexes however, it is essential to know their microstructures and stoichiometry since these properties will ultimately determine their range of applications [21].

In the present review, we focus on electrostatic complexation between charged-neutral diblock copolymers and oppositely charged nanocolloids such as surfactant micelles (section II) and magnetic nanoparticles (Section III). For both types of nanocolloids, core-shell hierarchical aggregates of sizes comprised between 20 and 500 nm were found under direct mixing conditions. In Section IV, we discuss alternative protocols for bringing oppositely charged species together. The association between charged-neutral copolymers and nanocolloids dispersed in brine solutions can be monitored by controlling the desalting kinetics. As a result, sphere- and rod-like aggregates with remarkable superparamagnetic and stability properties were obtained.

# II – Electrostatic Complexation between oppositely charged surfactants and copolymers

## II.1 – homopolyelectrolytes and surfactants

When homopolyelectrolytes are mixed to oppositely charged surfactants in aqueous solutions, two phenomena are successively observed. First, at very low polymer concentration the surfactants and polyelectrolytes co-assemble into aggregates similar to surfactant micelles. This phenomenon occurs at the critical aggregation concentration (cac) which is usually much lower than the regular critical micellar concentration (cmc) [54, 55, 50]. As the surfactant concentration is further increased, a phase separation is observed. The solution becomes turbid at the mixing of the two components, and after centrifugation it displays two separated phases. The bottom phase appears as a concentrated liquid phase or as a solid precipitate whereas the supernatant remains fluid and transparent. The conditions for the phase separation are that the ratio Z between the positive and negative charges is above a critical value $Z_C$, and that the salt content remains low [55]. Note that the adsorption of such electrostatic coacervate phases at liquid-solid interfaces were also reported. They revealed synergistic co-adsorption features and non equilibrium behaviors that can be exploited in the context of coating and anti-biofouling applications [56-58].

In recent years, increasing attention was placed on the structure of the concentrated solid phase because it is anticipated that the description of the precipitates is a prerequisite for the understanding of the complexation mechanisms. Using small-angle x-ray scattering techniques, it was shown that the solid precipitates exhibited liquid crystalline order reminiscent from the structures found in surfactant concentrated phases [46, 59, 47, 60-66, 48, 37]. Cubic, hexagonal and lamellar phases were identified, depending on if the surfactant self-assembled into spheres, cylinders and bilayers, respectively. In these crystalline mesophases, the polyelectrolyte chains are assumed to be adsorbed at the surface of the aggregates and to link them [47, 62, 48].

When the former ion-containing chain is now one part of a diblock copolymer, the other block being a neutral and water soluble chain, the phenomena described above are modified. Aqueous solutions of asymmetric diblocks with oppositely charged surfactants do not exhibit a phase separation. Instead, there is the formation of finite size colloids made from both surfactant and copolymers. Many different names were given to these aggregates. Colloids made with surfactants were dubbed block ionomer complexes [31], colloidal or electrostatic complexes [22, 34, 37, 39]. Co-assembled colloids made from oppositely charged block polypeptides or from proteins were called polyion complex micelles by Harada and coworkers [17, 18]. Those made solely from polymers were termed (inter)polyelectrolyte complexes [67] or complex coacervate core micelles [68, 4]. With surfactants, Kabanov and coworkers suggested the formation of a core-shell supramolecular assembly where the shell is made from the neutral blocks [31, 32, 36]. Using freeze fracture, these authors found that poly(ethylene oxide)-*b*-poly(sodium methacrylate) diblock copolymers and single tail surfactants arrange spontaneously into vesicles composed on closed bilayers from polymer bound surfactant arrays [69, 36].

## II.2 – Evidence of a Critical mixing ratio

In this part, we illustrate our approach of controlled co-assembly by considering the ternary system composed of an anionic-neutral block copolymer, a cationic surfactant and water. The polymer was a

poly(acrylic acid)-*b*-poly(acrylamide) with molecular weights 6500 g mol$^{-1}$ and 37000 g mol$^{-1}$ for each block respectively, abbreviated PANa$_{6.5K}$-*b*-PAM$_{37K}$ in the following [38]. The polymers were synthesized by Rhodia using the Madix technology for the formulation of personal care products [70]. The cationic surfactant, dodecyltrimethylammonium bromide (DTAB) is characterized by a C12 aliphatic chain and exhibits an hexagonal mesophase at high concentrations [71]. Its critical micellar concentration (cmc) in H$_2$O is 0.46 wt.% (15 mM) [72].

The mixing of oppositely charged species was controlled by pouring rapidly a surfactant solution into a stock polymer solution at the desired charge ratio. This protocol, also called *direct mixing* will be compared to other formulation techniques in Section IV. The relative amount of each component was monitored by the charge ratio Z, $Z = [S]/n_{PE}[P]$ where [S] and [P] are the molar surfactant and polymer concentrations and $n_{PE}$ (= 90) is the degree of polymerization of the polyelectrolyte block. The procedure consisting in mixing stock solutions at the same concentration was preferred because it allowed to explore a broad range of charge ratios and keep the concentration in the dilute regime, typically c < 1 wt.% [35, 37, 38]. The procedure also revealed the existence of a critical Z-value, noted $Z_C$ in the complex formation.

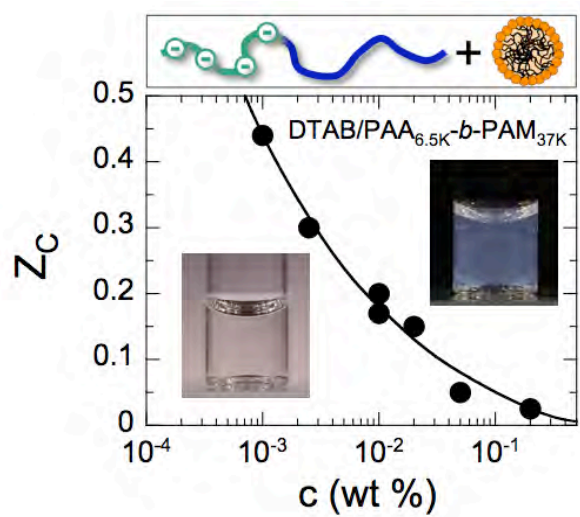

***Figure 1 :*** *Phase behavior of dispersions containing cationic surfactants (dodecyltrimethyl ammonium bromide, DTAB) and the anionic-neutral block copolymers (PANa$_{6.5K}$-b-PAM$_{37K}$) at pH 8 and T = 25 °C [22]. Above a critical mixing ratio $Z_C$ (close symbols), the dispersions were strongly light scattering, as illustrated by the upper right inset. The critical charge ratio $Z_C$ as a function of the total concentration c was determined from static light scattering. The continuous line resulted from best fit calculations using $Z_C(c) = 0.01 \times c^{-0.6}$. The panel above the figure illustrates the system considered in Section II for Figs. 1 to 5.*

Fig. 1 shows the critical charge ratio $Z_C$ as a function of the total concentration c. $Z_C$-values were derived from static light scattering experiments. They corresponded to the charge ratios where the scattering intensity exhibited a steep jump, increasing by 2 orders of magnitude [22, 37]. The results were obtained for c = 0.1 wt. % to 20 wt. % and showed that $Z_C$ decreases with increasing c as a power law : $Z_C(c) = 0.01 \times c^{-0.6}$ (continuous line in Fig. 1). These values coincided with the onset of phase separation found with the homopolyelectrolyte PAA$_{5K}$ [22]. In parallel, dynamic light scattering was performed as a function of Z and the hydrodynamic diameters were derived. Fig. 2 displays the evolution of the hydrodynamic diameter for c = 0.1 % and 1 wt. % as determined by dynamic light scattering using a Brookhaven spectrometer (BI-9000AT) operating at scattering angle θ = 90° and incident wave-length λ = 488 nm. In this figure, three scattering regimes were distinguished. At low Z, D$_H$ is 10 nm, a value consistent with the hydrodynamic diameter of a single diblock molecule [22]. Slightly above $Z_C$ (= 0.20), the autocorrelation function revealed the coexistence of two diffusive modes indicated by the dashed line : a fast mode associated to the diffusion of single diblock molecules, and a slow mode related to the diffusion of colloids of larger sizes. Between $Z_C$ and 1, $D_H$ passed through a maximum and stabilized at $62 \pm 8$ nm over a broad range of charge ratios. The colloids evidenced above $Z_C$ are those of interest in the present review.

To gain a deeper insight of their microstructure, cryo-transmission electron microscopy conducted out on a sample prepared again at isoelectric charge (Fig. 3) [35]. It revealed a existence of faintly contrasted spherical objects of average diameter 26 nm and polydispersity 0.16 (inset of Fig. 3). The value of 26 nm lied well below the actual hydrodynamic diameter measured by light scattering namely, $D_H = 62$ nm. This suggested that the colloids observed by cryo-TEM were actually surrounded by a polymer shell made from the neutral blocks of poly(acrylamide). Due to the low electronic contrast of poly(acrylamide) in water, the organic shell was not detected [35]. It is worth mentioning that the shell thickness $h$ derived here ($h = 17$ nm) was comparable to the hydrodynamic diameter of a single diblock ($D_H = 11$ nm).

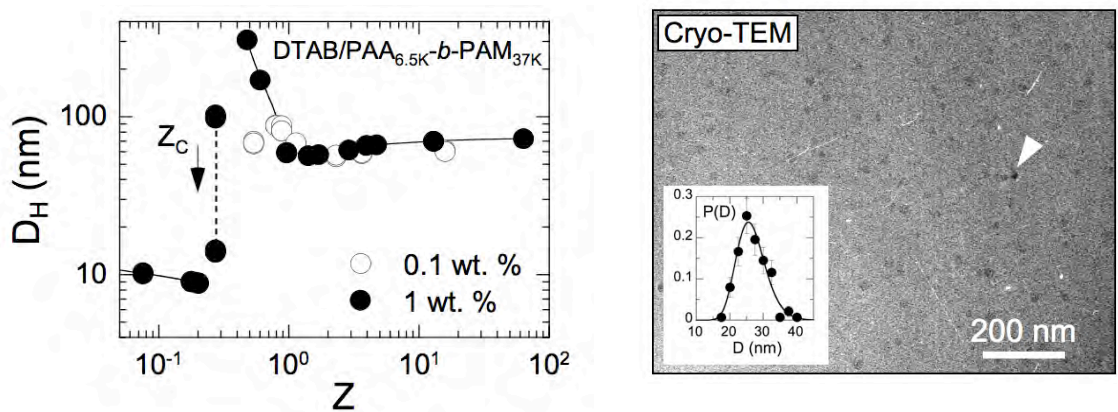

**Figure 2 (left) :** *Evolution of the hydrodynamic diameter $D_H$ as function of the charge ratio $Z$ for DTAB/PANa$_{6.5K}$-b-PAM$_{37K}$ mixed dispersions at $c = 0.1$ wt. % (open symbols) and at 1 wt. % (closed symbols) [22]. For $Z > 1$, $D_H$ reached a constant value around 62 nm. In this experiment, each data point represents a dispersion prepared by direct mixing (see text).*

**Figure 3 (right) :** *Cryo-transmission electron microscopy image of DTAB/PANa$_{6.5K}$-b-PAM$_{37K}$ electrostatic complexes prepared in D$_2$O at isoelectric charge ($Z = 1$) and concentration $c = 0.5$ wt. % [35]. Faintly contrasted objects (arrow) of average diameter 26 nm and polydispersity 0.16 are visible. Their size distribution is provided in the inset. As a result of an analysis based on light, neutron and x-ray scattering experiments, the objects seen by Cryo-TEM were ascribed to the cores of the electrostatic complexes illustrated in Fig. 5.*

## II.3 – Hierarchical microstructure revealed by neutron scattering

Because it allowed the quantitative determination of microstructures in the range $1 - 100$ nm, small-angle neutron scattering (SANS) was extensively used during the last decades [73]. Small-angle neutron scattering was performed at the Institute Laue-Langevin (Grenoble, France) on the beam lines D11 and D22 on samples prepared with D$_2$O for contrast reasons. Fig. 4 shows the neutron scattering intensity obtained for a DTAB/PANa$_{6.5K}$-b-PAM$_{37K}$ solution in D$_2$O at isoelectric charge (open symbols). Also included are the Rayleigh ratio data measured by static light scattering (closed symbols). Representative for this class of materials, the scattering intensity was characterized by three primary features :

1. A strong forward scattering as q → 0,
2. An oscillation at intermediate wave-vector and
3. A structure peak indicated by an arrow around $q^* = 0.165$ Å$^{-1}$.

As shown using a Porod representation [22, 34], the part of the scattering corresponding to Points **1** and **2** ($q < 0.05$ Å$^{-1}$) was identified unambiguously as arising from spherical colloids of average size 22 nm and polydispersity 0.16. According to this interpretation, the oscillation around $q = 0.04$ Å$^{-1}$ marked the location of the first minimum of the sphere form factor. The agreement between the cryo-TEM and SANS data for the determination of the typical colloidal sizes (26 nm *versus* 22 nm) was here excellent.

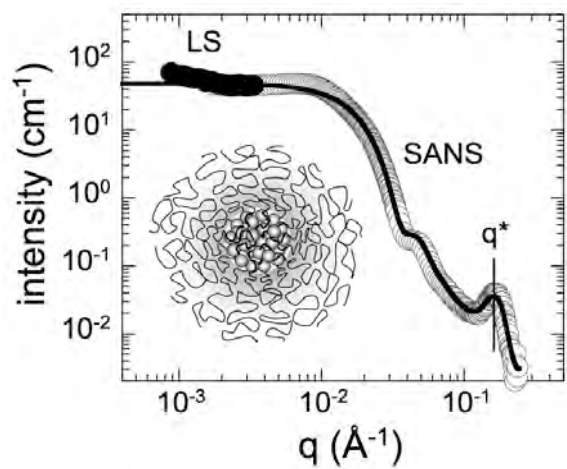

***Figure 4 :*** *Neutron scattering intensity obtained for a DTAB/PANa$_{6.5K}$-b-PAM$_{37K}$ solution in D$_2$O at isoelectric charge (open symbols). Static light scattering data are also included (closed symbols). The structure peak at q\* = 0.165 Å$^{-1}$ arises from a micro phase-separated state of strongly interacting micelles located in the core of the colloids. The continuous line results from Monte Carlo simulations using a model of colloid provided in the inset. Due to the low scattering contrast of the poly(acrylamide) in D$_2$O, the shell contribution was omitted in the simulations. Adjusting parameters are the number of micelles in the core (108) and the volume fraction (0.50) [38].*

Concerning the structure peak at large wave-vectors (Point **3**), several controls were performed to ascertain its origin. It was first shown to be independent on the concentration and on the charge ratio Z, and to be correlated with the strong forward scattering. Complementary experiments using deuterated surfactants or surfactants with longer aliphatic chains allowed us to draw the following conclusion : the structure peak at q\* arose from a concentrated phase of strongly interacting surfactant micelles located in the core of the aggregates. The fact that 2π/q\* ~ 4 nm, that is the diameter of a single DTAB micelle, reinforced the assumption of closely packed micelles. By analogy with the surfactant-polyelectrolyte coacervate phases, it was suggested that the micelles were linked together by the anionic blocks, the neutral segments being outside the core and dangling in the solvent. A representation of electrostatic complexes made from surfactants and copolymers is provided in the inset of Fig. 4 [22, 34]. This construct was said hierarchical because it involves two different length scales, the size of the single surfactant micelles (~ 4 nm) and that of the overall aggregate. This model was later on supported by Monte Carlo simulations [74, 35, 38, 75]. Simulations were employed to compute the scattered intensity arising from clusters of micelles. Due to the low scattering contrast of the poly(acrylamide) in D$_2$O [38], the contribution coming from the shell was neglected in the fitting. The adjusting parameters were the distribution of aggregation numbers, *i.e.* the number of spheres enclosed into the core and the volume fraction inside the core. The continuous line in Fig. 4 was obtained using an average of 108 micelles by core and a volume fraction of 0.50. The model accounted simultaneously for the three features mentioned previously (Points **1-3**).

In a later study, copolymers of different molecular weights and chemistry were investigated [37]. It was found that the core sizes of the complexes depended essentially on the structural charges borne by the charged block. In terms of stability, the optimal conditions were reached when the degree of polymerization of the neutral block was 2 - 5 times that of the charged block. Since our first SANS survey on the internal structure of surfactant-based electrostatic complexes, numerous colloidal systems with structure peaks at high wave-vectors were reported in the literature, and supported our view. This included surfactants [38, 76, 39, 77], proteins [78] or particles [75]. In conclusion, we show that oppositely charged surfactant and double hydrophilic block copolymers spontaneously co-assembled in the form of spherical hierarchical colloids, following a mechanism illustrated in Fig. 5. The question arises whether the above strategies could be generalized to other nanocolloids such as inorganic nanoparticles ?

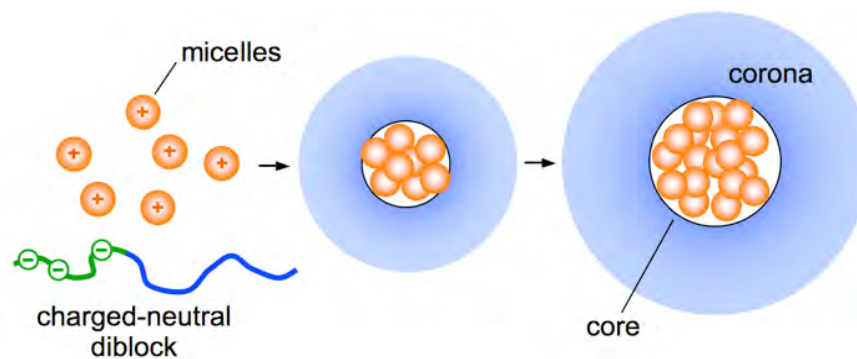

**Figure 5 :** *Schematic representation of the complex formation using oppositely charged surfactants and block copolymers.*

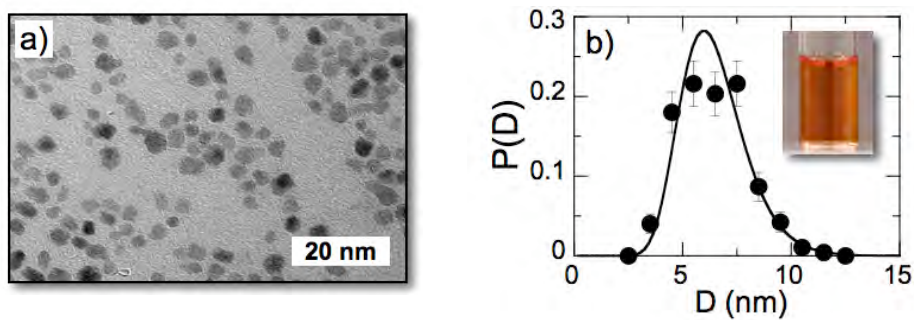

**Figure 6 :** *Upper panel : transmission electron microscopy (TEM) images of iron oxide nanoparticles investigated in this work, magnification ×120000. Lower panel : size distributions was derived from TEM images and fitted using a log-normal function, with median diameter 6.3 nm and polydispersity 0.23 [100]. The polydispersity was defined as the ratio between the standard deviation and the average. The inset shows a photograph of a vial containing a dilute iron oxide dispersion, c = 0.2 wt. %.*

# III – Generalization of the complexation protocols to inorganic nanoparticles

## III.1 – Existing Clustering strategies for inorganic nanoparticles

Inorganic nanoparticles made from noble metals, from oxides or from semiconductors are characterized by sizes comprised between 1 and 100 nm [79-81, 3]. In applications however, particulates with larger sizes may be required but yet remain difficult to generate by soft chemistry. Several issues become critical when the particle size increases, including colloidal stability, sedimentation and loss of physical properties. In order to circumvent these limitations, co-assembly strategies involving sub-10 nm nanoparticles were elaborated [82-92]. Nanoparticles were arranged into clusters of different dimensions and morphologies, with aggregation numbers ranging from few units to several thousands [82, 84, 87, 93, 89-92]. Because many nanomaterials are available in aqueous media, and because most of them display electric charges at their surfaces, the challenge was to generalize the approach developed for surfactants to inorganic particles. Concerning our work, sub-10 nm particles such as cerium oxide (CeO$_2$, nanoceria) [94-98], iron oxide ($\gamma$-Fe$_2$O$_3$, maghemite) [99-102, 95, 103, 104], europium-doped yttrium vanadate (Eu:YVO$_4$) [90] and yttrium hydroxyacetate (Y(OH)$_{1.7}$(CH$_3$COO)$_{1.3}$) [87, 105] were studied. As an illustration, we focus here on the model system made from magnetic nanoparticles ($\gamma$-Fe$_2$O$_3$).

## III.2 – Magnetic Nanoparticles as a model system

Iron oxide nanoparticles (bulk mass density $\rho = 5100$ kg m$^{-3}$) were synthesized according to the Massart technique [106, 107] by alkaline co-precipitation of iron(II) and iron(III) salts, oxidation of the magnetite (Fe$_3$O$_4$) into maghemite ($\gamma$-Fe$_2$O$_3$) nanoparticles, and by size-sorting by subsequent phase separations [108, 107]. At the end of the synthesis, the particles were dispersed in water at a weight concentration c ~ 10 wt. % and pH 1.8. At this pH, the particles were positively charged. The resulting electrostatic repulsion insured the colloidal stability of the dispersions over period longer than years. The magnetic fluids were characterized by electron microdiffraction, vibrating sample magnetometry, magnetic sedimentation and light scattering. For the dispersions investigated here, the size distribution as determined from TEM measurement (Fig. 6a) could be represented by a log-normal function, with median diameter $D_0^{NP} = 6.3$ nm and polydispersity $s^{NP} = 0.23$ (Fig. 6b) [100]. The particles were further functionalized with citrate ligands, a procedure which allowed to reverse the surface charges from cationic at low pH to anionic at high pH through the ionization of the carboxyl groups. For the citrate coated particles (Cit–$\gamma$-Fe$_2$O$_3$), the structural charge density was ascertained at -2$e$ nm$^{-2}$ by conductivity [109, 110] and light scattering measurements [90]. Cit–$\gamma$-Fe$_2$O$_3$ nanoparticles had thus sizes and surface charge densities that are larger than those of surfactant micelles (6.3 *versus* 4 nm for the diameter, -2$e$ *versus* +1$e$ nm$^{-2}$ for the density). Taking these figures into account, the inorganic particles had 5 times more structural charges at their surfaces as compared to surfactant aggregates (-250$e$ *versus* +53$e$) [90]. The sign of the electrostatic charges between the cationic micelles and the citrate-coated nanoparticles was also different, and cationic-neutral copolymers were used instead.

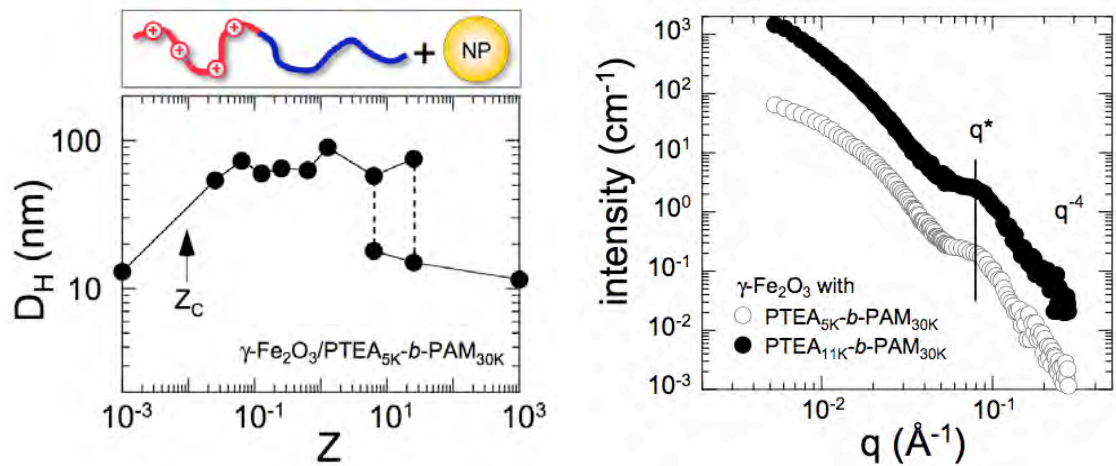

**Figure 7 (left) :** *Hydrodynamic diameter $D_H$ as function of the charge ratio Z for mixed dispersions made from citrate-coated iron oxide nanoparticles (Cit–$\gamma$-Fe$_2$O$_3$) complexed with cationic-neutral (PTEA$_{5K}$-b-PAM$_{30K}$) block copolymers [90]. With a charge density of -2e nm$^{-2}$, the structural charges of the particles was estimated at 250 elementary charges. The prominent features here are the existence of critical charge ratio $Z_C$ around $10^{-2}$ (arrow), a constant hydrodynamic diameter over more than 3 decades in Z, and the coexistence state between free and associated particles for Z > 10 (dashed lines). The panel above the figure illustrates the system considered in Section III for Figs. 7 to 9.*

**Figure 8 (right) :** *Scattering intensities for citrate-coated iron oxide nanoparticles complexed with PTEA$_{5K}$-b-PAM$_{30K}$ and with PTEA$_{11K}$-b-PAM$_{30K}$ block copolymers [95]. Experiments were carried out at c = 2 wt. % and in H$_2$O for contrast reasons. The intensity for Cit–$\gamma$-Fe$_2$O$_3$/PTEA$_{11K}$-b-PAM$_{30K}$ was shifted by a factor 10 for sake of clarity. The structure peaks observed at high wave-vector at q* = 0.084 Å$^{-1}$ were interpreted as the signature of dense micro-separated phases of particles.*

## III.3 – Iron oxide supracolloidal assembly

Iron oxide nanoparticles were complexed using poly(trimethylammonium ethylacrylate)-*b*-poly(acrylamide), a diblock that was synthesized by controlled radical polymerization [70, 111]. The molecular weights were 5 000 g mol$^{-1}$ and 11 000 g mol$^{-1}$ for the charged blocks and 30 000 g mol$^{-1}$ for the neutral chain. The polymers were abbreviated PTEA$_{5K}$-*b*-PAM$_{30K}$ and PTEA$_{11K}$-*b*-PAM$_{30K}$. Dynamic light scattering performed on Cit–$\gamma$-Fe$_2$O$_3$/PTEA$_{5K}$-*b*-PAM$_{30K}$ mixed solutions at c = 0.2 wt. % revealed the presence of one or two diffusive relaxation modes. Fig. 7 displays the evolution of the hydrodynamic diameters D$_H$ determined for these two modes as a function of the charge ratio. For Z above a critical value Z$_C$, the D$_H$'s were larger than those of the polymers and nanoparticles, and remained at a constant value of 65 ± 10 nm. For Z > 10, a second mode associated to the single nanoparticles became apparent.

The results of Fig. 7 bear a strong resemblance to the behavior of surfactant-based complexes. Both systems are characterized by the existence of a critical charge ratio Z$_C$ and by a constant diameter above Z$_C$. The findings of a constant hydrodynamic diameter over more than 3 decades in Z, as well as of coexistence between free and associated particles suggested moreover the existence of a fixed stoichiometry between the charges borne by the polymers and by the particles. Using a stoichiometric model to account for the light scattering results, it was found that the numbers of polymers per particle were 14 for PTEA$_{5K}$-*b*-PAM$_{30K}$ and 6 for PTEA$_{11K}$-*b*-PAM$_{30K}$ in the mixed aggregates [90].

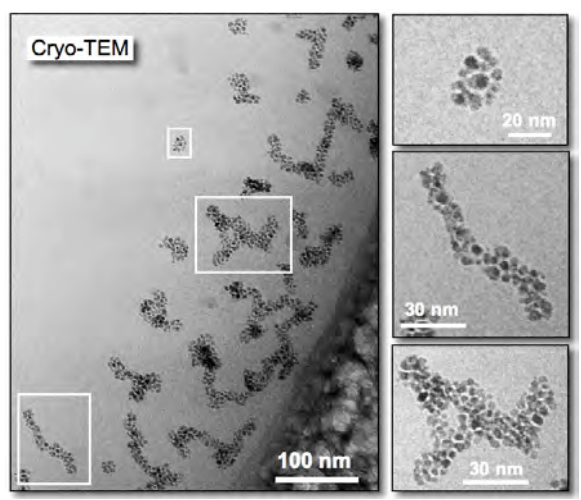

**Figure 9 :** *Cryogenic transmission electron microscopy (Cryo-TEM) images of nanoparticles clusters obtained by complexation between PTEA$_{11K}$-b-PAM$_{30K}$ and Cit–$\gamma$-Fe$_2$O$_3$. The aggregates were obtained according to the direct mixing technique. Their sizes vary from 20 to 200 nm [99, 95]. Inset : zoom of aggregates showing spherical, linear and branched morphologies.*

The analogy with the surfactant complexes was also evidenced using small-angle neutron scattering [95]. For SANS, organic-inorganic hybrids were prepared in H$_2$O because the scattering contrast was maximum. Fig. 8 compares the scattering intensities for aggregates obtained by complexation between Cit–$\gamma$-Fe$_2$O$_3$ particles with PTEA$_{5K}$-*b*-PAM$_{30K}$ and PTEA$_{11K}$-*b*-PAM$_{30K}$ block copolymers. Two of the three scattering features underscored for the surfactant-based complexes were again observed, namely the strong forward scattering (Point **1**) and the structure peak at high wave-vector (Point **3**). In Fig. 8, the structure peak was located at q* = 0.084 Å$^{-1}$, corresponding to an interparticle distance 6.6 nm and a volume fraction of 0.38 inside the clusters. These later values were derived from Reverse Monte Carlo simulations [95]. Note that because of the intrinsic polydispersity of $\gamma$-Fe$_2$O$_3$ particles, the expression 2$\pi$/q* could not be employed here to calculate the particle-particle distance in the clusters [112]. One characteristic feature was however missing for the Cit–$\gamma$-Fe$_2$O$_3$/PTEA-*b*-PAM hybrids, namely the oscillation at intermediate wave-vector. For the surfactants, the oscillation was ascribed to the narrow dispersity of the core size distribution. Transmission electron microscopy experiments were then conducted to confirm the SANS data, and thanks to the strong electronic contrast of iron, the entire aggregate morphology and inner structure were observed. Fig. 9 displays aggregates made by *direct mixing* of Cit–$\gamma$-Fe$_2$O$_3$ and PTEA$_{11K}$-*b*-PAM$_{30K}$. Covering a spatial field of $0.42{\times}0.56$ µm$^2$, the main frame in Fig. 9 displays well-disperse nanoparticle clusters. However, as anticipated from SANS, the aggregates were polydisperse in shape and in size. The right-hand side insets aim to illustrate the existence of spherical, linear and branched clusters, with dimensions varying from 20 nm to 200 nm [99, 95]. Similar findings were found out with non-magnetic 7 nm citrate-coated cerium oxide nanoparticles, indicating that the anisotropic structures of Fig. 9 did not stem from magnetic dipolar interactions [96, 98]. In most cases with inorganic nanoparticles [98], the same conclusions were reached. Even with an appropriate choice of polymer architecture, the *direct mixing* strategy was not satisfactory to control the formation of the co-assemblies, as it was the case for the surfactant micelles. Possible explanations for these behaviors are the large number of total charges involved in the process, or the fast kinetics of association between species [98].

# IV – The desalting transition : towards the fabrication of controlled size clusters and rods

## IV.1 – Alternative protocols for mixing species with opposite charges

The protocols for mixing oppositely charged species described in this section were inspired by molecular biology and were developed for the *in vitro* reconstitutions of chromatin. Chromatin stands for the DNA / histones macromolecular substance that forms the chromosomes of our cells [113]. The protocols applied here consisted first in the screening of the electrostatic interactions by bringing the dispersions of oppositely charged species to high salt concentration, and second in removing the salt progressively by *dialysis* or by *dilution* [114]. With this technique, the oppositely charged species were intimately mixed in solution but did not interact owing to the electrostatic screening. Compared to the *direct mixing* or *titration* methods introduced in the previous sections, these formulation processes revealed remarkable features such as the occurrence of an abrupt transition between a disperse and an aggregated state of particles.

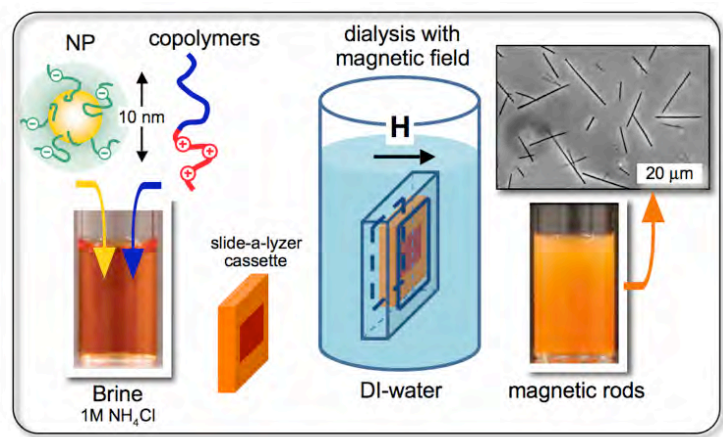

***Figure 10 :*** *Schematic representation of the protocol that controls the nanoparticle co-assembly and rod formation [101].*

The *dialysis* or *dilution* strategies involved in a first step the preparation of separate 1 M salted solutions containing respectively the anionic particles and the cationic-neutral diblock copolymers. In order to improve their stability, the $\gamma$-Fe$_2$O$_3$ particles were coated by poly(acrylic acid) with molecular weight 2000 g mol$^{-1}$ instead of citrate ligands as in the previous section [115, 100]. The thickness of the PAA$_{2K}$ brush tethered on the surface was estimated at 3 nm by dynamical light scattering. The salt used in the present work was ammonium chloride (NH$_4$Cl). The two solutions were then mixed with each other and it was verified by light scattering that the colloidal stability was not altered by the excess of salt. In a second step, the ionic strength of the dispersions was progressively diminished, by *dialysis* or by *dilution*. Dialysis was performed against deionized water using a Slide-a-Lyzer® cassette with molecular weight cut-off 10000 g mol$^{-1}$. The time evolution of the ionic strength was monitored by the measurement of the electric conductivity of the bath. Dialysis was carried under two different conditions, with or without magnetic field. The cartoon in Fig. 10 represents the case where the magnetic field (0.1 T) was on [116]. In the dilution process, deionized water was added stepwise to the nanoparticle and polymer salted dispersion, at a flow rate that was later translated into a desalting rate $dI_S/dt$, where $I_S$ denotes the ionic strength. With dilution, it was possible to vary $dI_S/dt$ from $10^{-5}$ to 1 M s$^{-1}$. With dialysis, the average rate of ionic strength change was of the order of $10^{-3}$ - $10^{-4}$ M s$^{-1}$ and the particle concentration remained practically constant.

## IV.2 – Evidence of a desalting transition as a function of ionic strength

Fig. 11 illustrates the light scattering results obtained by dilution using PAA$_{2K}$-coated γ-Fe$_2$O$_3$ nanoparticles and PTEA$_{11K}$-*b*-PAM$_{30K}$ copolymers, with c = 0.2 wt. % and charge ratio 1. Dynamic light scattering was monitored on a Brookhaven spectrometer (BI-9000AT) and on a NanoZS (Malvern Instrument) for measurements of the diffusion constant, and from which D$_H$ is derived [90]. The autocorrelation functions of the scattered light were interpreted using both cumulants and CONTIN procedure provided by the instrument software. After each step addition of de-ionized water, the scattering intensity and hydrodynamic diameter were measured. Doing so, the ionic strength was decreased at a rate $dI_S/dt \sim 10^{-4}$ M s$^{-1}$ comparable to that of dialysis. For the dispersion containing nanoparticles and copolymers (closed symbols), the data showed an abrupt upturn of the particle sizes at the critical ionic strength 0.39 M. This was not the case for the nanoparticles alone (open symbols), for which D$_H$ remained at the same level (20 nm). The transition between an unassociated and a clustered state was dubbed *desalting transition* because it occurred as the excess inorganic salt was removed from the dispersion [98, 116].

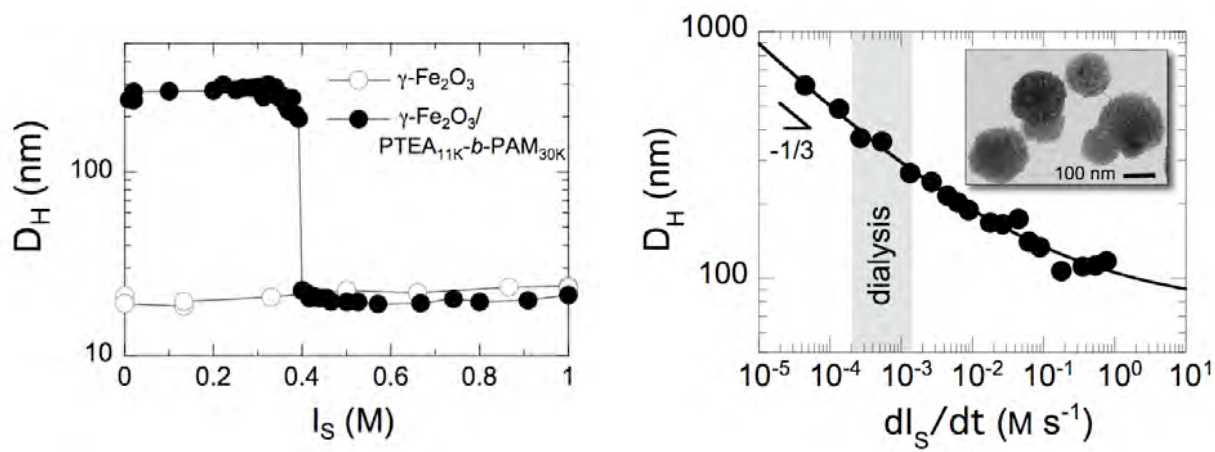

**Figure 11 (left)** : *Ionic strength $I_S$ dependence of the hydrodynamic diameter $D_H$ for PAA$_{2K}$–γ-Fe$_2$O$_3$ nanoparticles (empty symbols) and for a dispersion containing both PAA$_{2K}$–γ-Fe$_2$O$_3$ particles and PTEA$_{11K}$-b-PAM$_{30K}$ block copolymers (closed symbols). With decreasing $I_S$, an abrupt transition was observed at the critical value of 0.40 M [101]. The transition between the disperse and clustered state of nanoparticles resulted from the combined processes of electrostatic complexation of the cationic blocks with the particles and of slow nucleation and growth of aggregates.*
**Figure 12 (right)** : *Hydrodynamic diameters $D_H$ as a function of the desalting rate $dI_S/dt$ obtained via the dilution and dialysis protocols. Each data point represents a dispersion that underwent the desalting process illustrated in Fig. 11. At low desalting rates, $D_H$ displayed an asymptotic scaling of the form $D_H$*

~ $(dI_S/dt)^{-1/3}$. *Inset : TEM images of spherical aggregates obtained by dialysis. The aggregate size was estimated at 180 nm.*

In a second series of experiments, the desalting transition was monitored on a broad range of desalting rates, between $10^{-5}$ and 1 M s$^{-1}$. In Fig. 12, each data point represents the hydrodynamic diameter obtained from a dispersion that underwent this desalting process. Interestingly, the critical ionic strength did not depend on the desalting rate. For fast dilution, or quench ($dI_S/dt \sim 1$ M s$^{-1}$), the aggregates were found around 100 nm. The $D_H$-values in the fast dilution regime coincide well with those found by direct mixing, suggesting that in terms of kinetics, the mixing of oppositely charged species is equivalent to a quench [98]. With decreasing $dI_S/dt$, $D_H$ increased and displayed an asymptotic scaling law of the form $D_H \sim dI_S/dt^{-1/3}$. A power law with exponent -1/3 indicates that at infinitely slow dilution, the size of the clusters would diverge, the dispersion exhibiting then a macroscopic phase separation. This phase separation was observed by quenching the dispersion slightly below the critical ionic strength [98]. The inset of Fig. 12 displays a TEM image of aggregates obtained by dialysis. The aggregates of average diameter 180 nm were described as latex-type or composite colloids with a high load of magnetic particles. The volume fraction of magnetic materials was estimated around 0.25 inside the large spheres [95]. The temporal stability of the magnetic clusters was attributed to the fact that the structures formed were non-equilibrium structures.

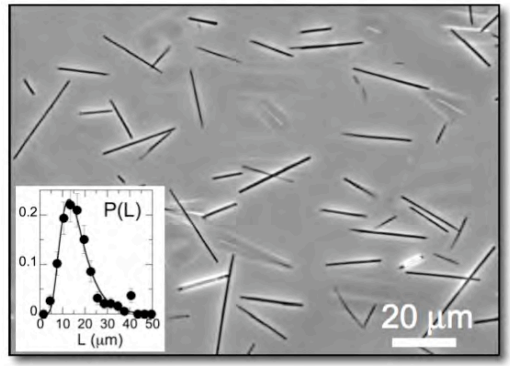

**Figure 13 :** *Phase contrast optical microscopy images (×40) of a nanorod dispersion [116]. Inset : Length distribution of rods obtained with 7.7 nm PAA$_{2K}$–γ-Fe$_2$O$_3$ particles. The continuous line was derived from best fit calculations using a log-normal distribution, with median length $L_0^{rod}$ = 15 µm and polydispersity $s^{rod}$ = 0.50. The polydispersity $s^{rod}$ was defined as the ratio between the standard deviation and the average.*

## IV.3 – Magnetic clusters and rods

In a third series of experiments, dialysis of mixed salted solutions was operated under a constant magnetic field of 0.1 T, *i.e.* corresponding to the set-up of Fig. 10. Once the ionic strength of the dialysis bath reached its stationary value, the magnetic field was removed and the solutions were studied by optical microscopy. Fig. 13 shows an optical transmission microscopy image of rods made of $D_0^{NP}$ = 7.7 nm γ-Fe$_2$O$_3$ particles at concentration c = 0.1 wt. % and charge ratio about 1 [101, 102, 104]. In the absence of magnetic field, anisotropic structures with random orientations and lengths comprised between 1 and 50 µm were clearly seen. Series of images similar to Fig. 13 were analyzed quantitatively to retrieve the rod length distribution. It was found to be well accounted for by a log-normal function with median length $L_0^{rod}$ (= 15 µm) and polydispersity $s^{rod}$ (= 0.5), as shown by the inset in Fig. 13. The nanostructured rods are defined as higher level objects because their construct involves three different length scales : *i)* the size of the elementary particles, around 10 nm, *ii)* the diameter of the rod which found between 200 and 500 nm and *iii)* the length of the rods comprised between 1 and 50 µm.

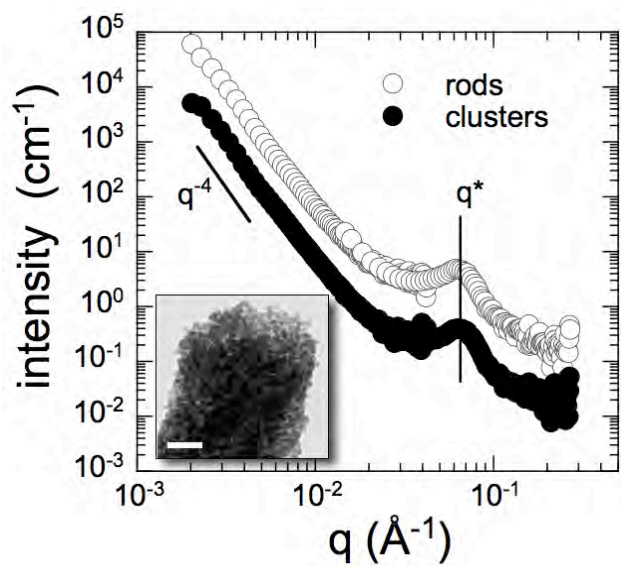

**Figure 14 :** *Neutron intensities of clusters ($D_H$ = 310 nm) and rods ($L_0^{rod}$ = 15 μm) obtained by electrostatic complexation between $PAA_{2K}–\gamma\text{-}Fe_2O_3$ nanoparticles and $PTEA_{11K}\text{-}b\text{-}PAM_{30K}$ block copolymers through the dialysis technique. The data for the rods were shifted by a factor 10 for clarity. The structure peaks observed at $q* = 0.064$ $Å^{-1}$ were interpreted as the signature of a dense packing of the particles within the supracolloidal objects. Inset : TEM image of the extremity of a rod [104].*

If a magnet was brought near to the sample, the rods reoriented spontaneously and followed the magnetic field lines. The coupling between the rods and the external field was shown to originate from the superparamagnetic properties of the elongated structures, a property that could be exploited in microfluidics and in therapeutics for the elaboration of micro-actuators [117]. Based on a comprehensive study of the morphology diagram of the rods, we elaborated on the mechanisms of rod formation. According to Ref. [116], the mechanism proceeds in two steps : first the formation and growth of spherical clusters of particles, and then the alignment of the clusters induced by the magnetic dipolar interactions. As far as the kinetics of the two processes is concerned, the clusters growth and their alignment occur concomitantly, leading to a continuous accretion of particles or small clusters, and a welding of the rodlike structure.

To finally demonstrate that the rods were nanostructured, small-angle neutron scattering was performed on dispersions prepared in $H_2O$, again for contrast reasons. The aggregates were fabricated by dialysis according to the protocol of Fig. 10 and using $PAA_{2K}–\gamma\text{-}Fe_2O_3$ nanoparticles of diameter $D_0^{NP}$ = 7.7 nm and polydispersity $s^{NP}$ = 0.22 [116]. The clusters had an hydrodynamic diameter of 310 nm and the rods a median length of $L_0^{rod}$ = 15 μm, with a polydispersity of 0.50. Fig. 14 shows that the scattering intensities for the two samples were very similar. The data for the rods were shifted by a factor 10 for clarity. From the three characteristics found for the surfactant-based complexes, only the structure peak at large wave-vectors, here located at $q* = 0.064$ $Å^{-1}$ remained (Point **3**). The two structure peaks are more intense than those of Fig. 8. This findings is in agreement with the fact that the scattering objects are now in the micrometer range. The low q-region exhibited typical Porod $q^{-4}$-dependences characteristic for large objects with abrupt interfaces. In the inset of Fig. 14, a TEM image of the extremity of a rod is illustrated, corroborating the dense packing of the particles in the cylindrical body. At the application of a constant magnetic field of strength 10 mT, the rods reoriented. Depending on their orientations with respect to the incoming beam, the scattered intensity collected on a two-dimensional detector remained isotropic or became strongly anisotropic (Fig. 15). In Fig. 15a, the magnetic field was parallel to the neutron beam and in the configuration of the spectrometer, the dispersion appeared isotropic to the incoming neutron [118, 119]. In contrast, when the field was in the plane of the detector as in Fig. 15b and 15c, the scattering intensity was enhanced in the reciprocal space in the directions perpendicular to the rods. In agreement with optical microscopy results, these findings confirmed that the basic structure of the rods was not altered by the application of a field, only their reorientations were modified.

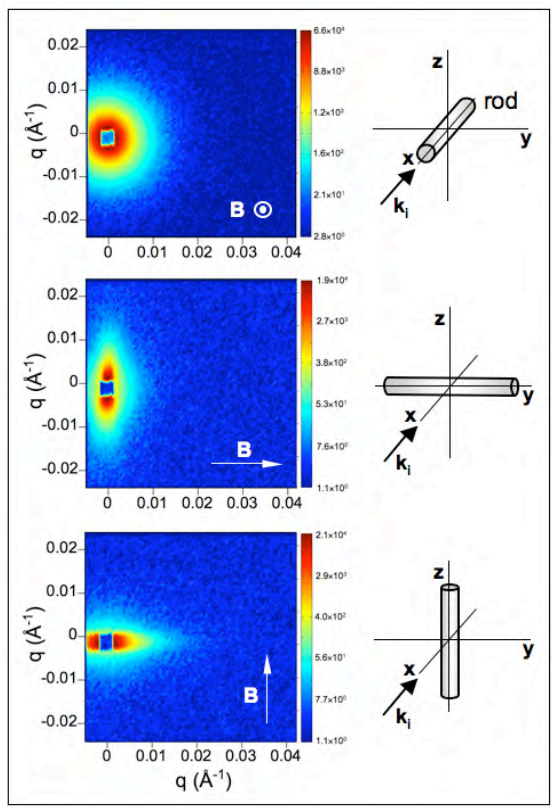

**Figure 15 :** *Two-dimensional scattering intensities obtained from iron oxide-based nanostructured rods. The nanorods were oriented by the application of a field of amplitude B = 10 mT : a) **B** // **k**$_i$; b and c) **B** ⊥ **k**$_i$, where **k**$_i$ denotes the wave-vector of the incident neutron beam.*

# V - Conclusion

In the preceding sections, we have shown that nanocolloids such as surfactant micelles and iron oxide magnetic nanoparticles share similar properties when mixed with ionic-neutral diblock copolymers : above a critical charge ratio, core-shell nanostructures form spontaneously under various mixing conditions. The driving forces for the formation of higher-level objects are insured by electrostatic attractions between the opposite charges, as well as by the release of the condensed counterions. In the following, we recapitulate the milestones that are important in the present experimental context, and suggest some applications for these nanostructured materials.

Our initial approach using diblock copolymers and surfactants was to design hierarchical aggregates of different sizes and morphologies by adjusting the architecture of the polymers. The goal was to assess the analogy between electrostatic complexes and polymeric micelles made from amphiphilic copolymers. For that, we studied synthetic copolymers with various architectures, including diblock, triblock, graft and star structures and of various molecular weights between 1000 to 100000 g mol$^{-1}$. The results of this strategy was very successful. On the two surfactant/copolymer systems investigated, DTAB/PAA-*b*-PAM and sodium dodecyl sulfate/PTEA-*b*-PAM [35, 37, 38], it was possible to produce aggregates of sizes lying between 20 and 500 nm. By changing the formulation conditions it was also possible to trigger morphologies going from spherical and cylindrical structures [77]. From these studies, it became clear however that electrostatic complexes depended on the preparation and mixing conditions, such as the pH, concentration, mixing speed etc.... We mentioned the case of DTAB/PANa$_{6.5K}$-$b$-PAM$_{37K}$ samples having been made at the same concentration, but using different pathways [22]. Their structures were still core-shell, but their neutron signatures indicated average cores of 22 and 47 nm respectively, with no evidence whatsoever of a late evolution of their internal structures. The same was observed with inorganic nanoparticles [99, 95]. These findings prompted us to search for alternative approaches, and in particular to look for processes where the complexation would be kinetically controlled instead of chemically controlled.

Many different techniques were already used to control the self- and co-assembly of macromolecular objects under strong attractive interactions. This includes confined impinging jets mixers [120] or microfluidic lab-on-a-chip devices [121]. In this work, we adopt a strategy borrowed from molecular biology for the *in vitro* preparation of chromatin. It is demonstrated that playing with the desalting rate allowed us to fine-tune the kinetics of association in a very accurate way. By varying the desalting rate from 1 to $10^{-5}$ M s$^{-1}$, nanoparticle clusters with average sizes 100 – 1000 nm were fabricated.

The enhanced stability of the nanostructures built up with surfactants and with inorganic nanoparticles is interpreted as resulting from a non-equilibrium associating process. Once the aggregates are formed, their microstructure remained unaltered over periods longer than years. The non-equilibrium character of these objects shows up with the nanorods made from sub-10 nm magnetic nanoparticles. The rods can be moved in the solvent or rotated by the application of a field or gradient without changing their internal structure, as shown by neutron scattering. Because the present strategy for electrostatic co-assembly of nanocolloids is simple and versatile, it should open new perspectives for the design of nanodevices, such as tips, tweezers, actuators etc., applicable in biophysics and biomedicine; for stimulating and sorting living cells; or as novel contrast agents for the drug delivery.

**Acknowledgement :** The present review would not have been possible without the extended network of colleagues who participated to this research. It is a pleasure to acknowledge here the collaborations I had over the years with Jean-Christophe Castaing, Jean-Paul Chapel, Galder Cristobal, Jérôme Fresnais, Bruno Frka-Petesic, Pascal Hervé, Eléna Ishow, Mikel Morvan, Julian Oberdisse, Régine Perzynski, Ling Qi, Olivier Sandre, Amit Sehgal, Minhao Yan and Kazuhiko Yokota. I would also like to thank Bernard Cabane, Valérie Cabuil, Andrejs Cebers, Martien Cohen-Stuart, Mathias Destarac, Eric Kaler, Christophe Lavelle, Sébastien Lecommandoux, Patrick Maestro, Nathalie Mignet, Lennard Picullel, Didier Roux, Serge Stoll for fruitful discussions. The Laboratoire Léon Brillouin in Saclay (Annie Brûlet, Fabrice Cousin) and the Institute Laue-Langevin in Grenoble (Isabelle Grillo, Ralph Schweins) are acknowledged for their technical and financial support. I would like to thank the Complex Fluids Laboratory (UMR Rhodia-CNRS n°166) and the Laboratoire Physico-chimie des Electrolytes, Colloïdes et Sciences Analytiques (UMR Université Pierre et Marie Curie-CNRS n° 7612) for our long-lasting collaborations and the later for providing us with the magnetic nanoparticles. During the last decade, this research work was supported in part by Rhodia (France), by the Agence Nationale de la Recherche under the contracts BLAN07-3_206866 and ANR-09-NANO-P200-36, by the University Denis Diderot Paris 7 (BQR2006), by the European Community through the project : "NANO3T—Biofunctionalized Metal and Magnetic Nanoparticles for Targeted Tumor Therapy", project number 214137 (FP7-NMP-2007-SMALL-1) and by the Région Ile-de-France in the DIM framework related to Health, Environnement and Toxicology (SEnT).